# Observation of orbital two-channel Kondo effect in a ferromagnetic $L1_0$-MnGa film

Lijun Zhu[1,2], Georg Woltersdorf[2] & Jianhua Zhao[1]

1. State Key Laboratory of Superlattices and Microstructures, Institute of Semiconductors, Chinese Academy of Sciences, P. O. Box 912, Beijing 100083, China.
2. Institut für Physik, Martin-Luther-Universität Halle-Wittenberg, von-Danckelmann-Platz 3, Halle 06120, Germany.

*Correspondence and requests for materials should be addressed to L.J.Z. (zhulijun0@gmail.com) and J.H.Z. (jhzhao@red.semi.ac.cn)

**The experimental existence and stability of the fixed point of the two-channel Kondo (2CK) effect displaying exotic non-Fermi liquid physics have been buried in persistent confusion despite the intensive theoretical and experimental efforts in past three decades. Here we report an experimental realization of the two-level system resonant scattering-induced orbital 2CK effect in a ferromagnetic $L1_0$-MnGa film, which is signified by a magnetic field-independent resistivity upturn that has a logarithmic and a square-root temperature dependence beyond and below the Kondo temperature of ~14.5 K, respectively. Our results not only evidence the robust existence of orbital 2CK effect even in the presence of strong magnetic fields and long-range ferromagnetic ordering, but also extend the scope of 2CK host materials from nonmagnetic nanoscale point contacts to diffusive conductors of disordered alloys.**

The overscreened Kondo effect displaying non-Fermi-liquid (NFL) physics has been of considerable scientific interest in recent years, especially due to their potential relevance to heavy fermions[1,2], topological superconductors[3], topological Kondo insulators[4], graphene[5], and quantum dots[6]. Its simplest manifestation, the two-channel Kondo (2CK) effect, may occur when a spin-1/2 impurity symmetrically couples to conduction electrons in two equal orbital channels via exchange interaction (spin 2CK)[6-8], or when a pseudospin-1/2 of two degenerate macroscopic charge states of a metallic island symmetrically couples to two conduction channels (charge 2CK)[9], or when a pseudospin-1/2 of structural two-level system (TLS, where an atom or atom group with small effective mass coherently tunnels between two nearby positions at a rate of $10^8$-$10^{12}$ s$^{-1}$) equally couples to two spin channels of conduction electrons via resonant scattering (orbital 2CK)[10-13]. The 2CK effect is expected to have a unique low temperature ($T$) resistivity upturn ($\Delta\rho_{xx}$), which scales with $\ln T$ beyond the Kondo temperature ($T_K$), followed by an exotic NFL behavior ($\Delta\rho_{xx}\sim T^{1/2}$) as the consequence of two conduction electron spins screening the spin (pseudospin) impurity[12-14]. The $T^{1/2}$ dependence of $\Delta\rho_{xx}$ is a hallmark of the NFL state in the 2CK effect, in striking contrast to the $T^2$ scaling of Fermi-liquid (FL) behavior in the case of the fully screened Kondo effect. Recently, the charge 2CK effect and spin 2CK effect were clearly demonstrated and channel asymmetry effect was probed directly and quantitatively[6,7,9]. However, the orbital 2CK physics has been under heated debate despite the intensive studies for almost 30 years. Even the sheer existence of the orbital 2CK effect is still controversial[15-23]. $T_K$ is given by ~exp(-1/4$JN(E_F)$), where $J$ and $N(E_F)$ are the exchange coupling strength and conduction-electron density of states at the Fermi energy ($E_F$), respectively. As shown by Aleiner *et al.*[15-17], for a TLS model that only considers a particle in a double well potential interacting with a degenerate electron gas, the orbital 2CK behavior can never be observed in the weak coupling limit ($JN(E_F) <<1$) because the energy splitting ($\Delta$) between the lowest two eigenstates of the TLSs always dominates the physics, i.e. $T_K < \Delta^2/T_K$, even if electron-assisted tunneling and the higher excitation states are taken into account[15]. However, as pointed by Zaránd[13], an experimental realization of the orbital 2CK effect with the generic low temperature resistivity upturn is expected for the TLSs with enhanced resonant scattering at $E_F$ and strong Kondo coupling ($JN(E_F)\sim1$), which is supported by the observation of the NFL behavior in ballistic conductors of Cu and Ti point contacts (PCs) fabricated by electron-beam lithography and diffusive conductors of ThAsSe glasses prepared by chemical vapor transport[18-24]. Furthermore, the stability of the orbital 2CK fixed point has remained an open question. A breakdown of the orbital 2CK fixed point is predicted at low energies $T_D$ (= $\Delta^2/T_K$) in the case of a nonzero $\Delta$ or asymmetric exchange coupling strength in the two channels[25,26]. Present theories also expect an imbalance in the channel population to quench the NFL behavior and to produce a crossover to FL behavior at a low $T$ in the neighborhood of $T$=0 2CK fixed point[26]. Experimentally, it has, however, remained unclear how robust the orbital 2CK fixed point is with respect to a channel asymmetry at finite temperatures. A magnetic field ($H$) of 5 T was reported to result in a breakdown of the NFL behavior at low energies in an early Cu PC experiment[24]. Some recent experiments argue for a negligible influence of strong magnetic field of up to 14 T in ThAsSe glasses[22,23] or even a slight spin polarization in $L1_0$-MnAl films[14,27], suggesting a considerable robustness of the orbital 2CK fixed points at finite temperatures. One reason was suggested to be that the electron spins are not directly involved in the Hamiltonian of the TLS coupling to the conduction electrons in the two spin channels[22]. It is, therefore, of great importance and interest to develop new Kondo systems with large $T_K$ and high-density TLSs in order to clarify the controversial physics of the orbital 2CK effect, especially its experimental existence and stability with respect to the population imbalance of two spin channels due to the strong magnetic fields or ferromagnetic exchange splitting.

The fully ordered $L1_0$-MnGa alloy is an itinerant magnet which is predicted to have a spin polarization of ~40% at the Fermi surface, a saturation magnetization ($M_s$) of ~2.51 $\mu_B$/Mn (i.e. 845 emu cm$^{-3}$), a ferromagnetic exchange splitting ($E_{exchange}$) of ~2.2 eV and a $E_F$ of ~11 eV (see Fig. 1a)[28,29]. Experimentally, $L1_0$-MnGa films with off-stoichiometry can be achieved in a wide Mn/Ga atomic ratio ($x$) range (0.76<$x$<1.75) by a non-equilibrium dynamic growth method, e.g. molecular-beam epitaxy[30]. Similar to the $L1_0$-ordered MnAl films[31,32], the magnetic and transport properties of $L1_0$-MnGa films are strongly dependent on the structural disorders and may be conveniently tailored by varying the growth parameters[33-35]. Therefore, $L1_0$-MnGa is an ideal playground for the exploration of disorder-related phenomena, e.g. orbital 2CK effect. In our previous paper[34], we observed in 50 nm thick disordered $L1_0$-Mn$_{1.5}$Ga films logarithmic low-$T$ resistivity upturns which exhibit a close relevance to growth temperatures ($T_s$) and an independence of strong applied magnetic fields. Here we show that the resistivity upturn in $L1_0$-MnGa films most likely arises from the orbital 2CK effect by taking an $L1_0$-MnGa film with enhanced disorder as an example. We observed a low-$T$ resistivity upturn with a clear transition from a ln$T$ dependence to NFL behavior signified by a $T^{1/2}$ dependence. The $T$ dependencies of the resistivity upturn are independent of applied magnetic fields up to 8 T. This result underpins the robustness of orbital 2CK effect even in the presence of strong magnetic fields and the spin polarization of the conduction electrons.

## Results

**Sample and ferromagnetism.** A 30 nm thick $L1_0$-MnGa film was grown on 150 nm GaAs-buffered semi-insulating GaAs (001) substrate at 200 ºC. The Mn/Ga atom ratio $x$ was determined by high-sensitivity x-ray photoelectron spectroscopy to be 0.94 (Fig. 1b). The chemical composition and the growth temperature were carefully chosen for an enhanced structural disorder. Figure 1c shows a cross-sectional inverse fast Fourier transform (IFFT) transmission electron microscopy (TEM) image of MnGa/GaAs interface, which clearly indicates the existence of the dislocations in the MnGa layer. The dislocations were suggested to be responsible for the TLSs[19,20,36]. Figures 2a and 2b show the well-defined perpendicular magnetization hysteresis loop and hysteretic Hall resistance measured at room temperature, respectively, revealing the ferromagnetism ($M_s$~100 emu cm$^{-3}$ at room temperature) and perpendicular magnetic anisotropy of this film. Figure 2c displays the $T$ dependence of magnetization ($M$) along film normal for the $L1_0$-MnGa film under $H$ = 50 Oe. The Curie temperature ($T_C$) of the film was determined to be 366 K following a three-dimensional (3D) Heisenberg model which expect $M \propto (T-T_C)^{1/3}$. The quick increase at temperatures below ~25 K is suggestive of nanoscale magnetic clusters embedded in the film due to its high degree structural disorders or due to the two sets of antiferromagnetically coupled Mn atoms which could have different magnetic moments and $T_C$ (similar to a ferrimagnet).

**Temperature dependence of the longitudinal resistivity.** Figure 2d shows the $T$ dependence of $\rho_{xx}$ for the $L1_0$-MnGa film at zero field ($H$ = 0 T) as an example. $\rho_{xx}$ shows a minimum at ~ 40 K, beyond which $\rho_{xx}$ increases monotonically with $T$ due to increasing thermal phonon and magnon scattering. Below this minimum, $\rho_{xx}$ shows an upturn down to 2 K which is the lowest $T$ that our present setup can reach. The same feature holds for different fixed $H$ of at least up to 8T. In the following we show that the low-$T$ resistivity upturn in our $L1_0$-MnGa film most likely arises from the TLS-induced orbital 2CK effect. In the absence of an external magnetic field, as displayed in Fig. 3a, $\rho_{xx}$ of the $L1_0$-MnGa film first varies linearly with ln$T$ below a temperature $T_0$ of ~25.5 K, similar to the well-known single-channel Kondo (1CK) effect due to static magnetic impurities. In fully screened 1CK systems[12], $\rho_{xx}$ was observed to saturate following the FL behavior (~$T^2$) at low $T$. Here, $\rho_{xx}$ deviates from the ln$T$ dependence and crossover to a $T^{1/2}$ dependence (Fig. 3b) when $T$ drops below $T_K$ of 14.5±1.5 K. Here the value of $T_K$ is defined as the center of the $T$-overlap between the two $T$ regimes of ln$T$ and $T^{1/2}$ dependences. The value of $T_K$ for the $L1_0$-MnGa film is comparable with that for the $L1_0$-MnAl film grown at 250 ºC (~13.5 K)[14] and ThAsSe glasses (~12 K)[22], but remarkably larger than that for metallic PCs (~5 K)[19,20]. The high value of $T_K$ suggests strong Kondo coupling between the TLSs and conduction electrons via resonant scattering, in the case of which present theories expect an experimentally accessible orbital 2CK effect.[13] The $T^{1/2}$-dependent resistivity is regarded as a unique signature of the NFL behavior for the 2CK effect[11,12,14]. Moreover, the temperature ranges of both the ln$T$ and $T^{1/2}$ behaviors are as wide as over 11 K in the $L1_0$-MnGa sample, which is comparable to or wider than those reported in other 2CK systems, e.g. ln$T$ ($T^{1/2}$) behavior only existed in the range of 2-6 K[18] (0.4-4 K[12]) for different Cu PCs. Here, we mention that an interpretation of weak localization or electron-electron interaction can be excluded. Firstly, both weak localization and electron-electron interaction (even if the diffusion channel is considered)[37] are in qualitative contradiction to the apparent transition from the ln$T$ scaling to the $T^{1/2}$ scaling at around $T_K$. Taking into account the resistivity of films ($\geq$ 148.4 $\mu\Omega$ cm), which yields a mean-free-path of ~11 nm at below 40 K, the film thickness of 30 nm, and the high crossover temperatures (that is, $T_K$) of ~14.5 K, a dimensional crossover seems impossible because neither the thermal length (relevant for electron-electron interaction) nor the inelastic scattering length (relevant for weak localization) is unlikely to approach the film thickness at this temperature. Furthermore, both weak localization and electron-electron interaction in a particle-particle channel are highly sensitive to magnetic fields[37], which is in obvious disagreement with the experiments.

The best fits to the data yield the slopes $\alpha$ = -$d\rho_{xx}/d(\ln T)$ ~1.38$\pm$0.02 $\mu\Omega$ cm ln$^{-1}$K for $T_K$ < $T$ < $T_0$ and $\beta$ = - $d\rho_{xx}/d(T^{1/2})$ ~0.71$\pm$0.02 $\mu\Omega$ cm K$^{-1/2}$ for 2 K< $T$ < $T_K$, respectively. For the strong coupling TLS centers in the diffusive transport regime[11], the volume density of the TLSs ($N_{TLS}$) can be estimated by $N_{TLS} \sim \frac{\Delta\rho_{xxm}}{\rho_{xx}} \frac{N(E_F)}{\tau_e}$, where $\Delta\rho_{xxm}$ and $\tau_e$ are the maximum resistivity upturn due to the TLSs and electron scattering time. Using $\Delta\rho_{xxm}/\rho_{xx}$~1.79%, $\tau_e$~10$^{-15}$ s, and $N(E_F)$~4 ×10$^{22}$ eV$^{-1}$cm$^{-3}$, $N_{TLS}$ can be estimated to be ~10$^{20}$ cm$^{-3}$, which yields an average distance of ~2 nm for different TLSs in the $L1_0$-MnGa film. It should be pointed out that the significantly enhanced $T_K$ and $N_{TLS}$ make the $L1_0$-MnGa film an advantageous TLS host material over the conventional nanoscale PCs[19,20]. Note that the breakdown energy scale of the NFL behavior ($T_D$) is below 2 K and has not been reached within our experimental setup, indicating a small $\Delta$ (<5.4 K) of the TLSs in this $L1_0$-MnGa film.

**Magnetic field effects.** Another characteristic of the TLS-induced orbital 2CK effect is the $H$ independence of the resistivity upturn. Magnetic fields should not have any observable influence on the resonant levels, coupling strength, and thus the effect amplitude via changing the population balance of the two spin channels of the conduction electrons because the Zeeman splitting is negligibly small (~ 0.9 meV at $H$ = 8T) in comparison to the width of energy band and $E_F$ of a host system (~10 eV), e.g. ferromagnetic $L1_0$-MnAl. In order to establish the orbital 2CK physics in $L1_0$-MnGa film, we further examined the effects of $H$ on both the ln$T$ and $T^{1/2}$ dependences of $\rho_{xx}$. As shown in Figs. 3a and 3b, the magnetic fields have no measurable influence on the $T$ dependence: $\rho_{xx}$ scales linearly with ln$T$ and $T^{1/2}$ at $T_K$ < $T$ < $T_0$ and 2 K< $T$ < $T_K$, respectively, under different magnetic fields ranging from 0 to 8 T. Note that, under perpendicular $H$, anisotropic MR and spin disorder scattering-induced MR should be negligible in a film with large perpendicular magnetic anisotropy, because of the orthogonal magnetization-current relation and the large energy gap in spin wave excitation spectrum. This is highly amenable to study the intrinsic $H$ dependence of a 2CK effect.

Figure 3c summarizes the values of the slopes $\alpha$ and $\beta$ as a function of $H$ for the $L1_0$-MnGa film. It is clear that both $\alpha$ and $\beta$ are independent of $H$, strongly suggesting a nonmagnetic origin of the resistivity upturn observed in $L1_0$-MnGa. Specifically, there is no measurable change in $T_K$ under different $H$ (Figs. 3a and 3b), suggesting a negligible effect of $H$ on the Kondo coupling strength, tunneling symmetry, and barrier height of the TLSs. However, our $L1_0$-MnGa epitaxial film does not show any sign of a breakdown of the NFL behavior due to a magnetic field of up to 8 T in the entire temperature range that is of interest, which suggests both a negligible influence of the applied magnetic fields on the population balance of the two spin channels and the robustness of the 2CK physics to a slight population imbalance. These observations provide strong evidence for the orbital 2CK effect being induced by TLSs originating from nonmagnetic impurities. A negative magnetoresistance (MR) is found to accompany the orbital 2CK effect in several different host systems[14,19,20,22,23]. Here, the $L1_0$-MnGa also shows a negative MR at high $H$ in the entire $T$ range (Figs. 4a-4c), which monotonically shrinks from -1.8% at room temperature to -0.5% at low temperatures and does not saturate even at $H$ = 8 T. As shown in Fig. 4b, the negative MR doesn't scale with $H^2$, which is in contrast the 1CK effect. The Bethe ansatz equations and conformal field theory for the 2CK problem predict an MR that scales with $H^{1/2}$ and ln$H$ in $T^{1/2}$ and ln$T$ regimes, respectively. Although the negative MR of the $L1_0$-MnGa film does scale linearly with $H^{1/2}$ (see Fig. 4c), it should be irrelevant to the orbital 2CK effect as indicated by the energy scale of the $H^{1/2}$ dependence (at least up to 300 K) and the absence of a ln$H$ scaling.

**Coexistence of the 2CK fixed point with ferromagnetism.** The evident coexistence of the 2CK physics and ferromagnetism is an intriguing observation. Although the two spin channels are still degenerate in energy because the Kondo coupling with a TLS is nonmagnetic and does not involve any spin variables, the population imbalance of the two spin channels due to the ferromagnetic exchange splitting of the conduction band could be significant in comparison to the magnetic field effects for the TLS model. In fully ordered $L1_0$-MnGa and $L1_0$-MnAl, the spin moments of Mn atoms are parallel due to ferromagnetic Ruderman–Kittel–Kasuya–Yosida interaction and the spin polarization is dominantly determined by the Mn $3d$ states[38]. In disordered samples, the Mn-Mn antiparallel alignment due to antiferromagnetic superexchange simultaneously reduces $M_s$ and spin polarization as a consequence of the cancelling contributions from the oppositely aligned Mn atoms [28,30,38,39]. For the disordered $L1_0$-MnGa film studied here, the value of $M_s$ is only 12.5% of the theoretical value for the fully ordered $L1_0$-MnGa, indicating a robust antiparallel alignment of the Mn magnetic moments and a very low degree of spin population imbalance. This could be the reason why the ferromagnetism does not quench the 2CK physics here. The robust 2CK effect observed in ferromagnetic systems, e.g. $L1_0$-MnGa and $L1_0$-MnAl, also hints that the fixed point of an orbital 2CK effect is more robust to the loss of spin population balance in comparison to that of a spin 2CK effect to the orbital channel asymmetry. However, a dilution of NFL behavior and an enhancement of $T_D$ due to the loss of spin population balance is expected in a ferromagnet with a partially spin-polarized conduction band[14]. It would be very interesting to quantitatively determine how the stability of 2CK fixed point varies with an enhancing population imbalance of the spin channels. More theoretical and experimental efforts are needed to better understand the exotic 2CK physics, especially in ferromagnetic hosts.

The TLSs play a central role in the orbital 2CK model, however, the identification of their microscopic nature is generally challenging. In nanoscale ballistic conductors of Cu and Ti PCs[19-21], the $1/f$ noises spectrum was introduced to hint the existence of dynamic motion of the atoms. However, it is difficult to separate the component of such $1/f$ noises arising from atomic motion in diffusive conductors[40], e.g. ThAsSe glasses, $L1_0$-MnAl and $L1_0$-MnGa films. A definite identification of the microscopic nature of the TLSs requires future developments of highly sensitive spectroscopic probing techniques. However, we noted that dislocation kinks seem to be responsible for the formation of TLSs as suggested by point contact experiments and theoretical calculations[11,19,20,36]. A logarithmic resistivity increase at low temperatures was also found to be associated with the increase in the dislocation density in different dilute Al alloys where dislocations were introduced by shock loading and extension at different temperatures[41]. Independent resonant scatter centers could also form along one single dislocation with large spatial extent[13]. Taking into account that the average distance of the dislocations (see Fig. 1c) appears to be of the same order with that of the adjacent TLSs (~ 2 nm), we, therefore, surmise that the nonmagnetic Ga atoms at the high-density dislocations likely play the role of TLS centers in present film. Here, the nonmagnetic nature of orbital 2CK effect and the significant disorder dependence rule out the possibility of the Mn atoms, the electrons nor the embedded magnetic clusters as TLS centers.

## Discussions

We have presented the experimental evidence which strongly suggests the occurrence of a robust orbital 2CK effect due to the electron scattering by high-density TLSs in ferromagnetic $L1_0$-MnGa film. The $H$-independent resistivity upturn scaling with $\ln T$ and $T^{1/2}$ in the two $T$ regimes below the resistance minimum are well consistent with the TLS model. The large $T_K$ of ~14.5 K suggests a strong Kondo coupling between the TLS centers and the surrounding conduction electrons via resonant scattering. The orbital 2CK effect in a ferromagnetic material points to a more robust fixed point for orbital 2CK with respect to the slight spin population imbalance in comparison to that for spin 2CK to the orbital channel asymmetry. Our observation also suggests that diffusive films of disordered ferromagnetic alloys, which were overlooked in the past 2CK physics studies, can be better TLS host materials than conventional nanoscale point contact devices fabricated by electron-beam lithography due to their high TLS densities (~$10^{20}$ cm$^{-3}$) and high Kondo temperatures. Our findings also imply that the nonmagnetic disordered alloys may also be potential host materials for realizing orbital 2CK physics because of the absence of the spin polarization that actually hurts the orbital 2CK effect. This greatly extends the scope of TLS host systems for future studies of the orbital 2CK physics, which should be inspiring for future 2CK physics studies. More experimental and theoretical efforts are needed in the future in order to better understand the intriguing robustness of the 2CK physics even in the presence of ferromagnetism.

## Methods

**Sample preparation and characterizations.** The sample was prepared by a VG-80 molecular-beam epitaxy system with two growth chambers (one for growing III-V group semiconductors, the other for growing magnetic alloys). A semi-insulating GaAs (001) substrate was first loaded into the semiconductor chamber to remove the oxidized surface by heating up to 580 °C in Arsenic atmosphere (~1×10$^{-7}$ mbar) and to get a smooth fresh surface by growing a 150 nm GaAs buffer layer. Afterwards, the sample was transferred to second growth chamber to grow a 30 nm thick $L1_0$-MnGa film at 200 °C and a 5 nm thick MgO capping layer for protection from oxidation. The composition, structure, and magnetism were measured by an x-ray photoelectron spectroscopy (Thermo Scientific ESCALAB 250Xi), a transmission electron microscopy (JEOL 2010), and a Quantum Design superconducting quantum interference device (SQUID-5) magnetometer, respectively.

**Device fabrication and transport measurement.** The film was patterned into 60 μm wide Hall bars with an adjacent electrode distance of 200 μm using standard photolithography and ion-beam etching for transport measurements. The Hall resistivity ($\rho_{xy}$) and longitudinal resistivity ($\rho_{xx}$) were measured in a Quantum Design physical property measurement system (PPMS-9) as a function of temperature and magnetic field with a 10 μA excitation current, respectively. The applied magnetic fields were orthogonal to the film plane to minimize possible magnetoresistance effects.

**Acknowledgments**

We thank R. A. Buhrman for helpful discussion, T. P. Ma and S. S. P. Parkin for use of PPMS, S. H. Nie, D. Pan, J. X. Xiao and J. Lu for help during sample growth. We acknowledge the support by NSFC (Grand No. 61334006), MOST of China (Grant No. 2015CB921503), and priority program (SPP 1538) of the German research foundation (DFG).

**Author contributions**

L. J. Z. designed and performed the experiments, L. J. Z., G. W., and J. H. Z. analyzed the data and wrote the manuscript.

**Additional Information**

**Competing financial interests:** The authors declare no competing financial interests.

**Figures**

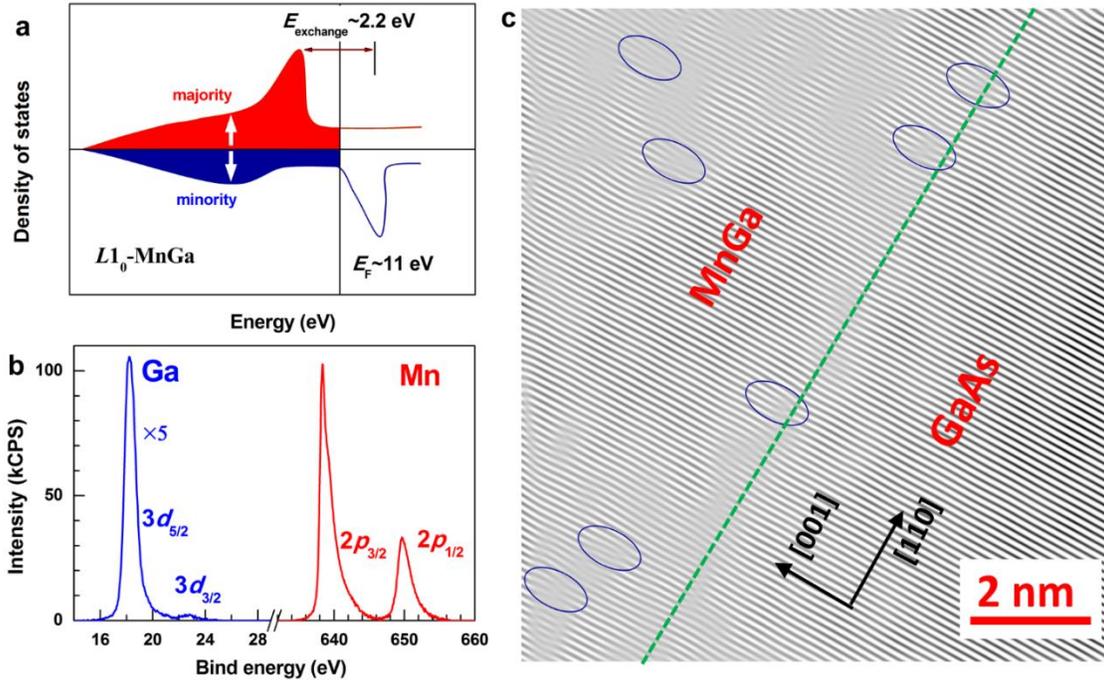

Figure 1. **Electronic and crystalline structure**. (a) Schematic depiction of partial density of states of fully ordered $L1_0$-MnGa; (b) x-ray photoelectron spectrum and (c) cross-sectional IFFT-TEM image of the $L1_0$-MnGa film. The green dashed line marks the GaAs/MnGa interface. The blue elliptic circles guide the position of a portion of the dislocations. The two black arrows refer to [001] and [110] crystalline directions of GaAs, respectively.

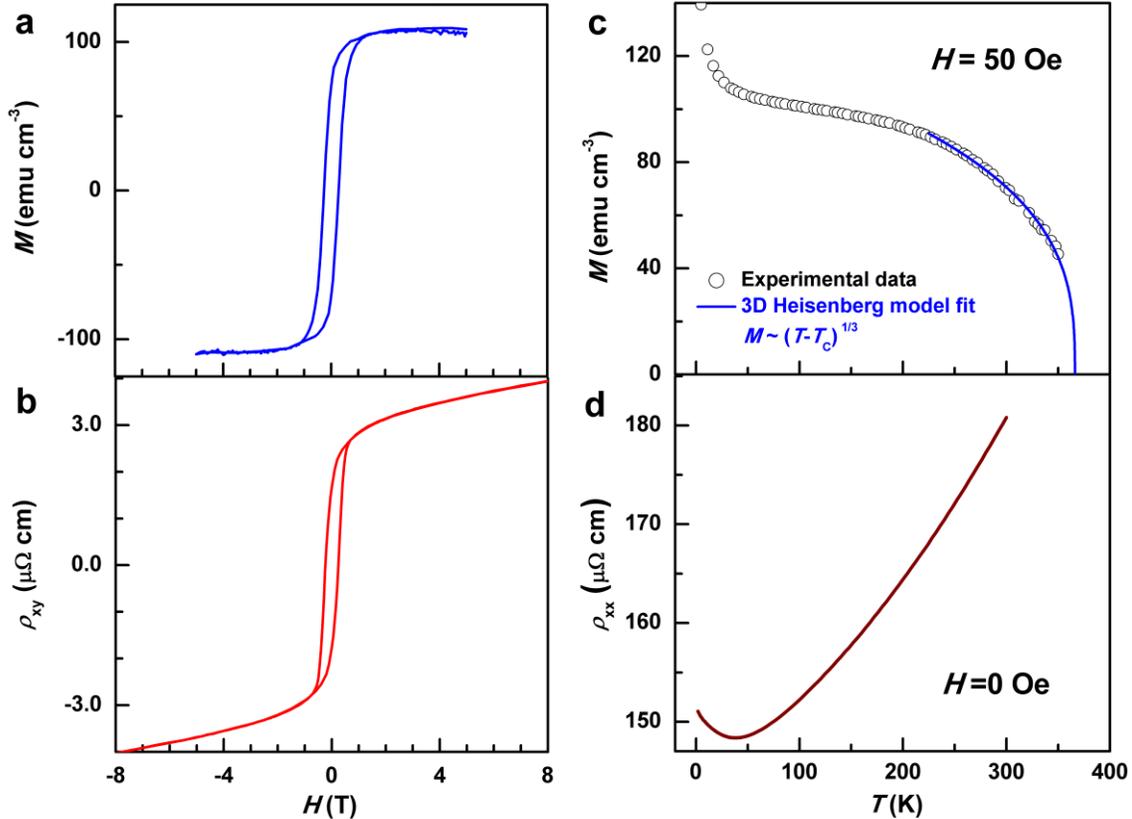

Figure 2. **Magnetic and transport properties.** (a) Magnetization hysteresis and (b) hysteretic Hall resistivity ($\rho_{xy}$) at 300 K, (c) $M$-$T$ curve, and (d) $\rho_{xx}$-$T$ curve at zero field for the $L1_0$-MnGa film. The blue solid line in (c) refers to the extrapolation of the $M$-$T$ data following the 3D Heisenberg model, from which the Curie temperature ($T_C$) was determined to be 366 K.

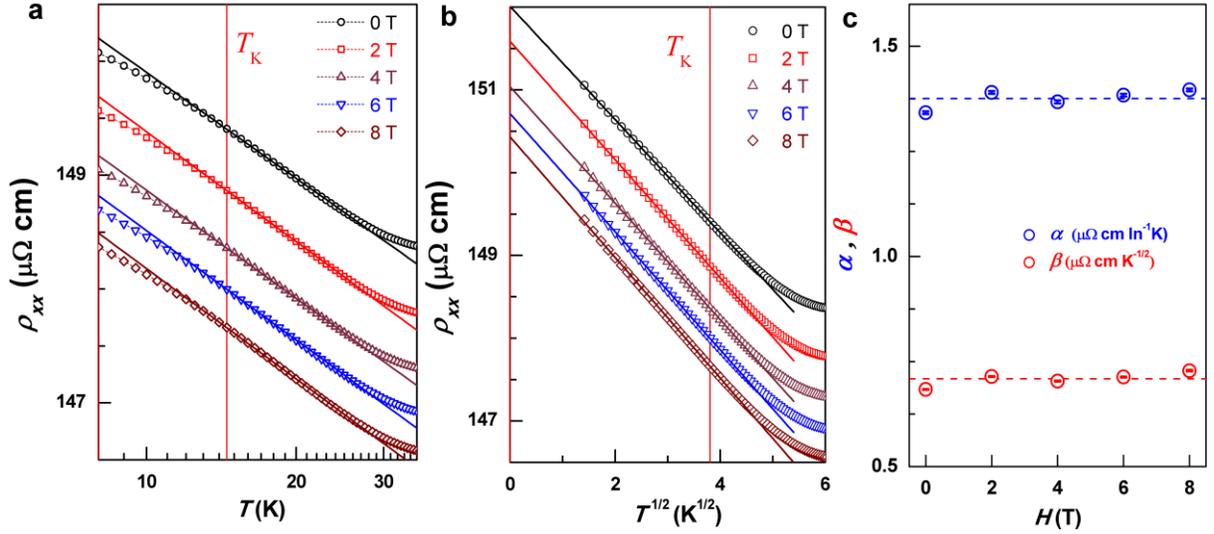

Figure 3. **Low-temperature resistivity upturn under different perpendicular magnetic fields.** (a) Semilog plot of $\rho_{xx}$ versus $T$ and (b) $\rho_{xx}$ versus $T^{1/2}$; (c) $H$ dependence of $\alpha$ and $\beta$ for the $L1_0$-MnGa film. For clarity, $\rho_{xx}$ was shifted by 0, -0.2, -0.4, -0.6, and -0.8 $\mu\Omega$ cm in (a) and (b), respectively. The Kondo temperature $T_K$ is 14.5±1.5 K. The dashed lines in (c) are for eye guidance.

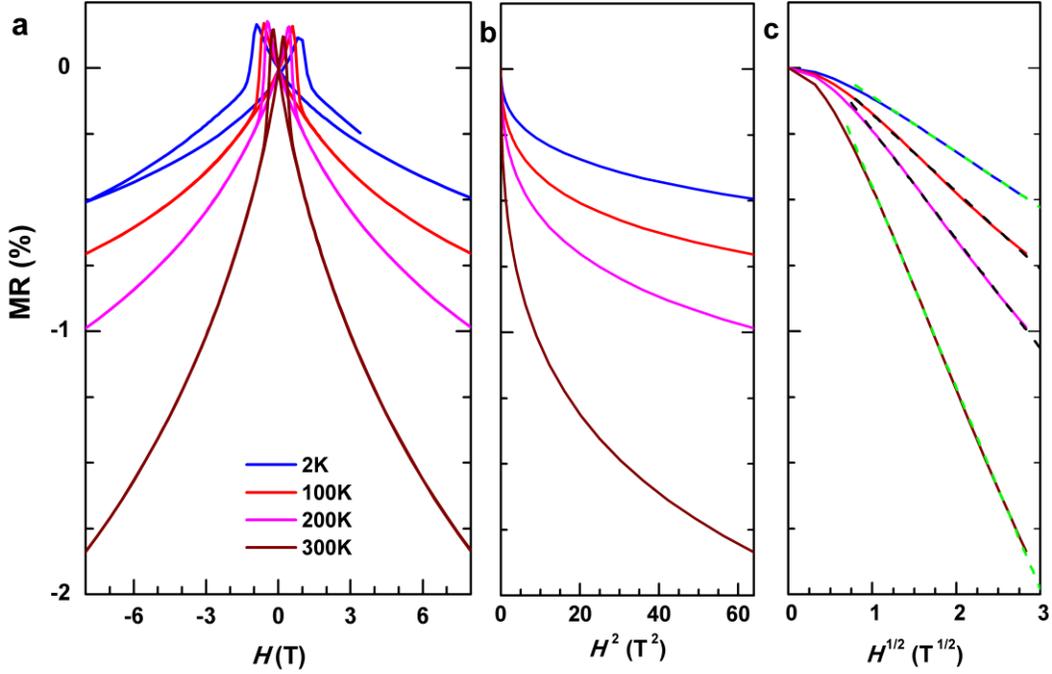

Figure 4. **Magnetoresistance**. (a) MR versus $H$, (b) MR versus $H^2$, and (c) MR versus $H^{1/2}$, respectively. The dashed lines in (c) represent the best linear fits of MR-$H^{1/2}$ for each temperature.